# Hexagonal-boron nitride substrates for electroburnt graphene nanojunctions

Hatef Sadeghi, Sara Sangtarash and Colin Lambert

Quantum Technology Centre, Department of Physics, Lancaster University, Lancaster LA1 4YB, UK

*h.sadeghi@lancaster.ac.uk; c.lambert@lancaster.ac.uk*

**Abstract**

We examine the effect of a hexagonal boron nitride (hBN) substrate on electron transport through graphene nanojunctions just before gap formation. Junctions in vacuum and on hBN are formed using classical molecular dynamics to create initial structures, followed by relaxation using density functional theory. We find that the hBN only slightly reduces the current through the junctions at low biases. Furthermore due to quantum interference at the last moments of breaking, the current though a single carbon filament spanning the gap is found to be higher than the current through two filaments spanning the gap in parallel. This feature is present both in the presence of absence of hBN.

**Introduction**

It is a pleasure to write this short paper in memory of Marcus Büttiker. The Lambert group has used Landauer-Büttiker formulae for more than 30 years, starting with disordered systems in the early 1980's [1, 2] when the validity of such formulae was hotly debated. Since that time the group has applied these formulae to a range of problems in Andreev scattering [3], phonon transport [4, 5], tunnelling through single molecules and chains [6, 7], molecular spintronics [8] and most recently though nanopores leading to new strategies for DNA sequencing [9, 10]. This paper is the latest in this long line and uses the Landauer-Büttiker formula to evaluate the electronic properties of electroburnt graphene nanoelectrodes.

The high thermal and mechanical stability of graphene, combined with its zero band gap and two-dimensional lattice [11], make it an ideal material for use as nano-electrodes [10, 12-

14]. Recently electroburnt graphene nanojunctions [14] have attracted increasing scientific interest, because their sub-nanometer gaps are sufficiently small to be spanned by single molecules. Indeed molecules with planar anchor groups bind particularly strongly to the surface of the graphene via π-π and van der Waals interactions and form stable electrode-molecule-electrode junctions [15]. Electroburnt graphene nanaojunctions can be grown on a silicon oxide substrate with a buried gate electrode, which provides a versatile three terminal platform for exploring and tuning electron transport through single molecules. However in such junctions, the thickness of the oxide barrier between the gate and molecule means that the electrostatic coupling of the gate electrode to the molecule is inefficient. Therefore strategies to increase the coupling are needed.

In this paper, our aim is to study the properties of graphene electroburnt junctions formed on an insulating hexagonal-boron nitride (hBN) substrate, which would allow a gate electrode beneath the hBN layer to be located much closer to the graphene electrode gap, thereby increasing the gating efficiency. Even in the absence of a molecule, electroburnt graphene junctions on silicon oxide or free-standing in vacuum exhibit unexpected quantum interference effects at the last moment of burning, just before the gap forms. In this paper, our aim is to examine electron transport through graphene nanojunctions and to determine if such effects persist in junctions formed on hBN.

Using the method described in [14], we used the molecular dynamics code *LAMMPS* [16] to generate 42 different initial examples of graphene electrodes near the moment of breaking. In each case the electrodes were covalently connected by a series of randomly generated carbon filaments or a constricted neck of graphene. Geometry relaxation was then carried out for each structure using the density-functional-theory (DFT) code *SIESTA* [17] with the same parameters as described in [14]. Next the transmission probability $T(E)$ of the electrons with energy $E$ passing through the junction was calculated using our *GOLLUM* transport code [18] and finally the Landauer-Büttiker formula [19], was used to compute the electrical conductance and current-voltage relation for each structure.

**Result and discussion**

Figure 1 shows an example of a nanojunction on free standing graphene (fig. 1a) and on a hBN substrate (fig. 1b). After geometry relaxation, the graphene is approximately A-B stacked on the hBN with a slight lattice mismatch [20]. The separation between the two layers is calculated to be approximately *3.4Å* in agreement with reported experimental [21] and theoretical values [22]. The substrate size is chosen to be larger than graphene to avoid the effect of hBN edges on the current. The corresponding transmission coefficient and current through the device is shown in figures 1c and 1d respectively. The transmission coefficient around the Fermi energy is almost the same in the absence and presence of hBN although at higher energies new resonances due to the weak coupling between carbon and boron or nitrogen atoms appear. Due to the fact that the transmission is only slightly changed around the neutrality point, the low-bias current is barely affected by the hBN. At higher voltages, the presence of additional transmission resonances around $E_F = -0.8V$ could cause the current in the presence of hBN to increase.

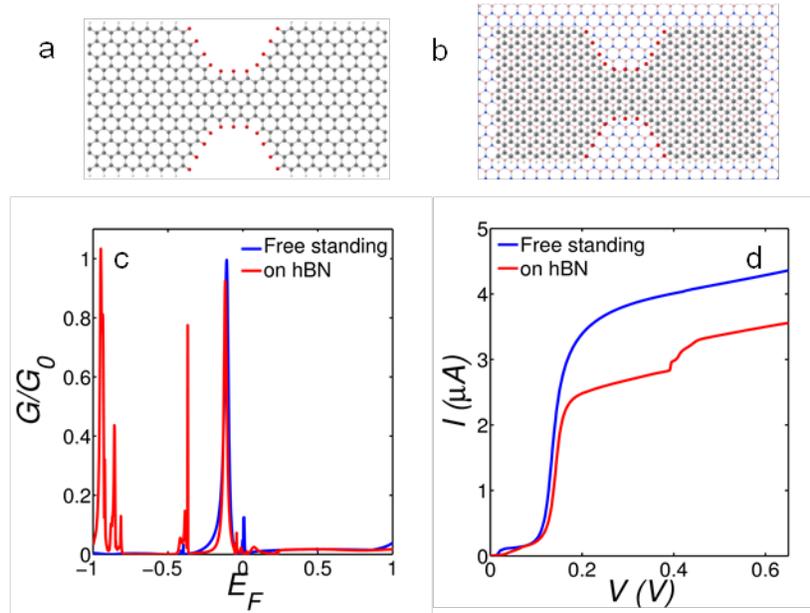

Figure 1. Graphene nano-junctions (a) free standing graphene, (b) on hexagonal boron nitride substrate, (c) the conductance in different Fermi energies and (d) current – voltage characteristic.

To create a junction with nano-meter size gap, a sufficiently-high bias voltage is applied to burn a pre-patterned constriction of graphene [14]. When the graphene begins to burn, feedback control is activated and the bias voltage is dropped. This process is repeated several times to break the junction. One complicating feature is the presence of the oxygen atoms around the burning site, which can affect the shape and size of the junctions. This oxygen could be supplied from air or in vacuum from the $SiO_2$ oxide substrate. The use of hBN could avoid this feature and help to form more reproducible junctions. Figure 2 shows five examples of such a junctions created in the absence of the oxygen atoms. Structures 3, 4 and 5 are formed from successively narrower constrictions, which are stable after DFT relaxation. In contrast, a constriction formed from a single chain of hexagons without edge termination is not stable and forms two parallel carbon chains, as shown in s structure 2. The most narrow structure 1 is formed from a single carbon chain and has relaxed to form a series of single and triple bonds in agreement with the known properties of oligoynes [23-25]. Such chains have been recently observed experimentally using TEM on free standing graphene [26].

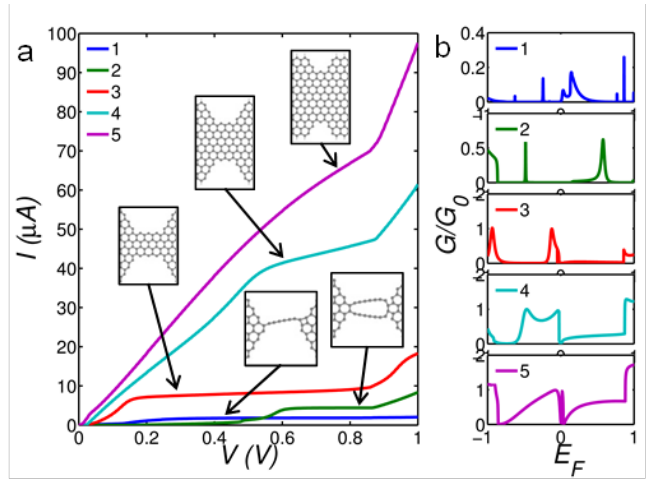

Figure 2. The junctions without termination formed in vacuum condition in the absence of the oxygen either from the air or the oxide substrate. (a) current – voltage characteristic of the junctions 1-5, (b) the conductance of each junction in different Fermi energies.

The conductances of devices 1 to 5 are shown in figure 2b. Junctions 1 and 2 of figure 2a are interesting, because the low-bias current (ie at voltages less than 0.5V) of the single carbon chain (device 1) is higher than device 2 (two parallel carbon chains) in low bias voltages for a wide bias window. This is highly non-classical, since one would expect a higher conductance for two parallel resistors than for a single resistor. In our recent paper [14], we discussed the origin of this counterintuitive phenomenon and demonstrated that it is a signature of room-temperature quantum interference and arises from a combination of the semimetallic band structure of graphene and a cross-over from electrodes with multiple- path connectivity to single-path connectivity just before breaking. In molecular scale objects such room-temperature quantum effects also play important role [27, 28].

We placed all of the 42 different free standing graphene junctions discussed in ref. [14], on hBN substrates and relaxed them using DFT. Examples of the relaxed structures are shown in figure 3a (the substrate not shown to provide clear picture of the junction, in all junctions the hBN is present under graphene junction similar to what shown in figure 1b). For the structure of figure 3a, figure 3b shows the current of each configuration in different bias voltage with and without hBN substrate. In general in the presence of the hBN, the current slightly drops due to the weak coupling between the graphene-hBN double layers. However, it is shown that the current trough the graphene is normally higher on the hBN substrate rather than $SiO_2$ [29]. The structures 1, 2, 4, 5 and 6, where a broken junction, a junction with multiple paths or a junction with very weak coupling is formed, yield in general lower currents than the structure 3, where a single path junction formed. This trend remains unchanged even in the presence of the hBN substrate although the magnitude of the current in all cases decreased in the presence of the hBN substrate. This decrease in the current in the presence of hBN occurs for all of the 42 different configurations.

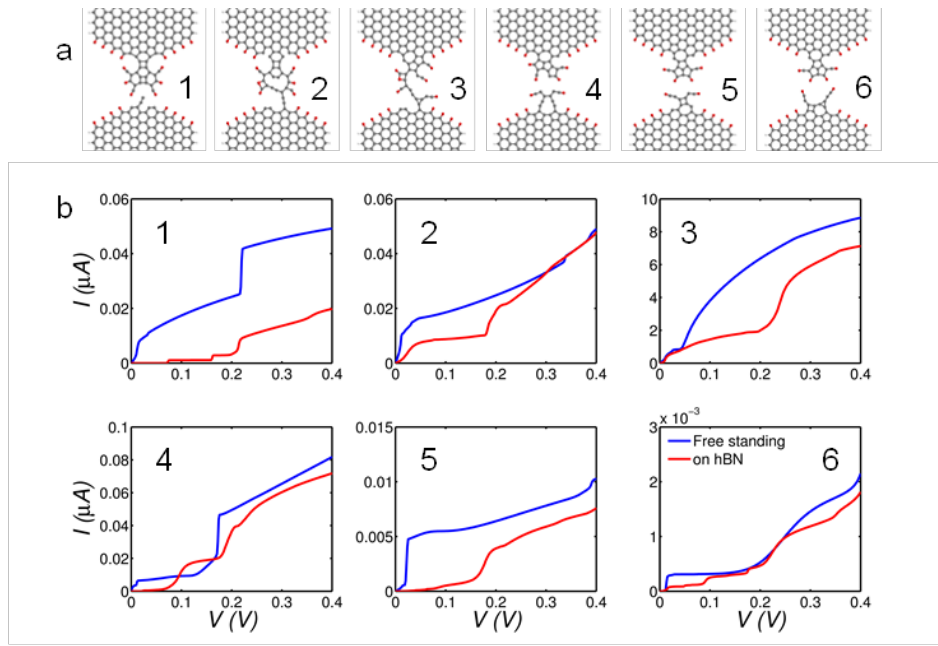

Figure 3. Six examples out of 82 different configurations studied in this paper. ($a_{1-6}$) Graphene nano-junctions with different carbon filament in the junction, ($b_{1-6}$) current – voltage characteristic of the junction shown in (a) without and with a hBN substrate.

As noted in [14], for free-standing graphene nanojunctions, the current in the device 3 is expected to be higher than the current in the other devices due to the quantum interference effect, because destructive interference occurs in multipath junctions. Figure 3 confirms that this feature persists in the presence of hBN.

## Conclusion

We have compared current-voltage characteristics of graphene nanojunctions in the absence and presence of a hBN substrate and found that typically the current is decreased by the presence of hBN. We have also found that the current flow in a single-filament junction is higher than that of a double-filament junction, due to the quantum interference, both in the presence and absence of hBN. These results demonstrate that electron transport through electroburnt graphene junctions are only slightly perturbed by the presence of a hBN substrate. Since the latter allows a gate electrode to be placed in close proximity to the electrode gap, this suggests that graphene nanogaps on hBN substrates are a viable route to high-efficiency gated devices.

## Acknowledgment

This work was supported by the European Commission (EC) FP7 ITN "MOLESCO" project no. 606728 and UK EPSRC, (grant nos. EP/K001507/1, EP/J014753/1, EP/H035818/1).